# Effective mobility for sequential carrier transport in multiple quantum well structures

Kasidit Toprasertpong,[1,*] Stephen M. Goodnick,[2] Yoshiaki Nakano,[1] and Masakazu Sugiyama[1]

[1]School of Engineering, the University of Tokyo, Bunkyo-ku, Tokyo 113-0032, Japan
[2]School of Electrical, Computer and Energy Engineering, Arizona State University, Tempe, Arizona 85287, USA
***Corresponding author**, e-mail: toprasertpong@hotaka.t.u-tokyo.ac.jp

## ABSTRACT

We investigate a theoretical model for effective carrier mobility to comprehensively describe the behavior of the perpendicular carrier transport across multiple quantum well (MQW) structures under applied electric field. The analytical expressions of effective mobilities for thermionic emission, direct tunneling, and thermally-assisted tunneling are derived based on the quasi-thermal equilibrium approximation and the semi-classical approach. Effective electron and hole mobilities in InGaAs/GaAsP MQWs predicted from our model are in good agreement with the experimental results obtained from the carrier time-of-flight measurement near room temperature. With this concept, the complicated carrier dynamics inside MQWs can be simplified to an effective mobility, an equivalent parameter which is more straightforward to handle and can be easily incorporated in the conventional drift-diffusion model.

## I. INTRODUCTION

Due to the unique optical and electrical properties of low-dimensional materials, multiple quantum well (multiple QW, MQW) structures have found numerous applications in optoelectronics. The employment of MQWs is, however, usually accompanied by the poor perpendicular carrier transport, which limits the bandwidth in high-speed devices [1,2], the carrier injection efficiency in light-emitting devices [3,4], and the photocarrier collection efficiency in photovoltaic devices [5,6]. For decades, a great deal of theoretical and experimental studies have been carried out to gain our understanding on the carrier dynamics in such structures, including carrier scattering, capture, thermalization, and escape processes [7-13]. In practice, the actual carrier dynamics is the result of simultaneous combination of all the processes. The complexity of the resultant dynamics limits its investigation to the numerical self-consistent calculation on a large equation set of related processes [14-17]. Moreover, unless the special treatment is implemented, the incorporation of these processes is incompatible with the conventional drift-diffusion model, which is widely used as a tool for the device output estimation and the design optimization. The need for specific simulators and the lack of simple models cause a large gap between the fundamental study and the device application.

To avoid handling the periodic potential profile in the drift-diffusion analysis, some research works have simplified the complicated carrier transport in nanostructures inside the devices with some carrier mobilities. It is often assumed similar mobilities to the bulk materials [18-20], despite the fact that there are reports observing much lower mobility values [21-23]. This is due to insufficient theoretical models that can conveniently estimate the mobility of perpendicular carrier transport from the structure/material parameters.

The theoretical investigation on mobility so far has been mainly focused on the transport in strongly coupled MQWs, or so-called superlattices, where carrier transport is dominated by the Bloch transport through the miniband. The Esaki-Tsu mobility [24], which is derived from the kinetics in the momentum space of the miniband, is a well-known expression for the Bloch transport, and there are several attempts to modify the expression for the mobility [25,26]. On the other hand, less studies have been investigating the mobility in MQWs which have weak or no coupling. In this case, dynamics is dominated by the sequential (hopping) transport: carriers move from QW to QW. Furthermore, even in strongly coupled superlattices, the transition of the dominant transport from the Bloch to the sequential type is expected if the scattering rate is high enough to break the coherency of the carrier wavefunction [27,28], for instance, at high temperature. D. Calecki et al. [29] have proposed the hopping mobility for the photon-limited process, but only the numerical calculation is available due to the complicated interaction between carriers and phonons. There are several other literatures investigating the perpendicular sequential transport across MQWs [11,30,31], but only little attention has been paid to the transport at the mobility (linear transport) regime.

In this study, we investigate the carrier effective mobility in periodic MQW structures close to room temperature. In Section II, we derive the formulae for the effective mobility corresponding to the sequential transport through the thermionic emission, sequential direct tunneling, and thermally-assisted tunneling processes. Our aim is to formulate explicit expressions to minimize the additional numerical calculations such as the scattering by various types of phonons, and make them reachable to a wide range of applicable areas; therefore we have employed some appropriate approximations including the quasi-thermal equilibrium and the semi-classical approach, instead of the rigorous treatment of quantum mechanics. We also consider the modification of the Schneider-Klitzing thermionic emission model [11], which assumes wide QWs with negligible quantum-confinement effect, as pointed out in [32], and neglects the perturbation of the distribution function by the electric field. In Section III, the effective mobility estimated by our model is compared with the experimental results measured from the time-of-flight technique. We discuss about the possibility to simplify MQWs as quasi-bulks with the effective mobility in the drift-diffusion analysis in Section IV and summarize our findings in Section IV.

## II. MODEL

### A. Sequential transport

We consider a periodic MQW structure with the well thickness $w$, the barrier thickness $b$, and the period $L = w + b$. At high temperature, e.g. room temperature, the scattering rate (< 1 ps) [7,8,14] is high enough that carriers are rapidly captured and thermalized as they cross any QW. This implies that the one-step process model can well describe the carrier dynamics in the MQW system. That is, carriers in the $j$th QW can escape to the adjacent QWs following the rate equation

$$\frac{dn_j}{dt} = \frac{n_{j-1}}{\tau_{\text{esc}+}} - \frac{n_j}{\tau_{\text{esc}+}} - \frac{n_j}{\tau_{\text{esc}-}} + \frac{n_{j+1}}{\tau_{\text{esc}-}} - \frac{n_j}{\tau_r}, \quad (1)$$

where $n_j$ is the carrier density in the $j$th QW, $1/\tau_r$ is the carrier recombination rate, $1/\tau_{\text{esc}+}$ and $1/\tau_{\text{esc}-}$ are the carrier escape rates when moving forward and backward, respectively. By using the Fokker-Planck approximation $n_{j\pm 1} \approx n_j \pm \partial n_j/\partial j + \partial^2 n_j/\partial j^2$ and $z = jL$, we obtain a solution $n_j(t)$ with the center moving at the velocity [11,33]

$$v = L\left(\frac{1}{\tau_{\text{esc}+}} - \frac{1}{\tau_{\text{esc}-}}\right), \quad (2)$$

which corresponds to the average carrier velocity in the MQW. In a symmetric structure, the escape process is symmetric ($\tau_{\text{esc}+} = \tau_{\text{esc}-}$) and hence there is no net velocity. The transport symmetry begins to collapse when the electric field $F$ is applied. If we assume the electric field strength is in the linear regime, we can write the escape rate as $1/\tau_{\text{esc}\pm} = 1/\tau_{\text{esc}}|_{F=0} \pm \left(\partial(1/\tau_{\text{esc}})/\partial F\right)_{\text{asym}}\big|_{F=0} F + \left(\partial(1/\tau_{\text{esc}})/\partial F\right)_{\text{sym}}\big|_{F=0} F$, where $1/\tau_{\text{esc}}$ is the general expression for the carrier escape rate across one side of barriers. The second term corresponds to the asymmetric field perturbation (e.g. the shift of barrier bandedge position) and the third term to the symmetric perturbation (field direction-independent perturbation, e.g. the energy level shift due to the quantum-confined Stark effect). In this way, the low-field average velocity can be expressed by

$$v = 2LF \frac{\partial}{\partial F}\left(\frac{1}{\tau_{\text{esc}}}\right)_{\text{asym}}\bigg|_{F=0}. \quad (3)$$

By using the analogy of the carrier mobility, it is reasonable to define a parameter $\mu_{\text{eff}}$ by

$$\mu_{\text{eff}} = 2L \frac{\partial}{\partial F}\left(\frac{1}{\tau_{\text{esc}}}\right)_{\text{asym}}\bigg|_{F=0}. \quad (4)$$

We call this the *effective mobility* since most of carriers in the MQW are not free carriers but still, macroscopically, behave similarly to the drift transport. It should be noted that the symmetric term $\left(\partial(1/\tau_{\text{esc}})/\partial F\right)_{\text{sym}}$ and the higher expansion terms in the escape rate will result in the field-dependent effective mobility, but this effect will be small in the low-field regime [33]. We can formulate the effective mobility for any escape process using Eq. (4) if the dependency of the escape rate on the electric field is known. We consider that carriers can escape out of a QW through three processes: the escape of carriers with energy higher than the energy barrier, the tunneling process for carriers in the ground state, and the tunneling process for carriers distributed in the higher quantum-confinement states. These processes are also known as the thermionic emission (the thermal escape), the direct tunneling, and the thermally-assisted tunneling, respectively.

### B. Thermionic emission process

We firstly investigate the thermal escape rate through one side of the energy barrier within the effective mass approximation. In the following, we refer to the conduction band when considering electrons and the valence band for holes. Fig. 1(a) shows the bandedge diagram of the QW with the unperturbed bandedge energy $E_0$ and the barriers with the band-offset $qV_b$. If the carrier thermalization is sufficiently fast, carriers in the same QW and band are always in the thermal equilibrium and it is possible to define the quasi-Fermi level $E_F$ in each QW. Then, the distribution function $f_{3D}$ of carriers in the continuum levels in the absence of electric field is

$$f_{3D} = \frac{1}{4\pi^3} e^{-\frac{\hbar^2 k^2 / 2m^*_{3D} + E_0 - E_F}{k_B T}}, \tag{5}$$

where $\hbar = h/2\pi$ is the reduced Planck constant, $k_B$ is the Boltzmann constant, $T$ is the temperature, and $m^*_{3D}$ is the effective mass of carriers in the continuum levels. Under electric field $F$ applied in the $z$ direction, similarly to the perturbation treatment in the Boltzmann transport equation [34], the distribution function is perturbed by

$$\Delta f_{3D} = -\frac{q\tau_{sct} F}{\hbar} \frac{\partial f_{3D}}{\partial k_z}, \tag{6}$$

where $q$ is the elementary charge and $\tau_{sct}$ is the carrier scattering time.

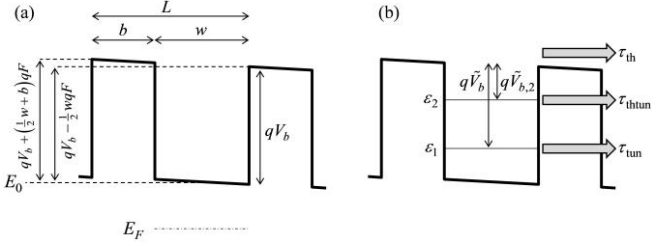

FIG. 1. Schematic bandedge diagrams of a QW and notations: (a) Structure and band parameters, (b) Effective barrier heights and escape processes. Similar notations are used for the conduction, heavy-hole valence, and light-hole valence bands.

The carrier total energy of $E_0 + qV_b + \frac{1}{2}(b \mp L)qF$ is required to escape thermally to the next QW in the forward and backward direction, respectively. In other words, those carriers must have the $z$ component of the wavevector $k$ more than $k_{b\pm} \equiv \sqrt{2m^*_{3D} q V'_{b\pm}/\hbar^2}$, where $V'_{b\pm} \equiv V_b + \frac{1}{2}(b \mp L)F$. Then, the current density $J_{th\pm}$ flowing out of one side of the QW due to the thermal velocity is given by

$$J_{th\pm} = q \int_{k_{b\pm}}^{\infty} \int_{-\infty}^{\infty} \int_{-\infty}^{\infty} \frac{\hbar k_z}{m^*_{3D}} (f_{3D} \pm \Delta f_{3D}) dk_x dk_y dk_z$$

$$= \frac{q\hbar}{4\pi^3 m^*_{3D}} \int_{k_{b\pm}}^{\infty} \int_{-\infty}^{\infty} \int_{-\infty}^{\infty} k_z \left(1 \pm \frac{q\tau_{sct} \hbar F}{m^*_{3D} k_B T} k_z\right) \tag{7}$$

$$e^{-\frac{\hbar^2 k^2 / 2m^*_{3D} + E_0 - E_F}{k_B T}} dk_x dk_y dk_z.$$

The upper sign is for the escape toward the forward direction and the lower sign for the backward direction. (We use this plus-minus notation in the derivation thereafter.)

The total carrier density $n$ in the QW layer for the non-degenerate condition can be obtained by summing the carrier densities in the confined states $n_{2D,i}$ and the continuum states $n_{3D}$

$$n = \sum_i n_{2D,i} + n_{3D}$$

$$= \sum_i \frac{m^*_{\|i} k_B T}{w\pi \hbar^2} e^{-\frac{E_0 + \varepsilon_i - E_F}{k_B T}}$$

$$+ \int_{|k_z| > k_{b0}} \int_{-\infty}^{\infty} \int_{-\infty}^{\infty} f_{3D} dk_x dk_y dk_z, \tag{8}$$

where $\varepsilon_i$ ($< qV_b$) is the $i$th confinement energy relative to $E_0$, $m^*_{\|i}$ is the in-plane effective mass, and $k_{b0} \equiv \sqrt{2m^*_{3D} qV_b/\hbar^2}$. For $qV_b$, $\varepsilon_{i+1} - \varepsilon_i > 2k_B T$, which satisfy in most short-period MQW systems, it is a good approximation to take

$$\exp[qV_b/k_B T] \gg 1, \tag{9a}$$

$$n \approx n_{2D,1} = \frac{m^*_{\|1} k_B T}{w\pi \hbar^2} e^{-\frac{E_0 + \varepsilon_1 - E_F}{k_B T}}. \tag{9b}$$

Under the situation (9a), $J_{th\pm}$ in Eq. (7) becomes

$$J_{th\pm} \approx \frac{qm^*_{3D} k_B^2 T^2}{2\pi^2 \hbar^3} e^{-\frac{E_0 - E_F}{k_B T}} e^{-\frac{\hbar^2 k_{b\pm}^2}{2m^*_{3D} k_B T}} \left\{1 \pm \frac{q\langle \tau_{sct} \rangle \hbar k_{b\pm} F}{m^*_{3D} k_B T}\right\}$$

$$= \frac{qm^*_{3D} k_B^2 T^2}{2\pi^2 \hbar^3} e^{-\frac{E_0 - E_F + qV'_{b\pm}}{k_B T}} \left\{1 \pm \sqrt{2m^*_{3D} qV'_{b\pm}} \frac{\mu_{3D} F}{k_B T}\right\}. \tag{10}$$

Here, $\int_x^{\infty} \xi^2 e^{-\xi^2} d\xi \approx \frac{1}{2} x e^{-x^2}$ for $x \gg 1$ (see Appendix A), is used. $\mu_{3D} \equiv q\langle \tau_{sct} \rangle / m^*_{3D}$ is the so-called drift mobility, limited by the scattering process of the free carriers. Then, the thermal escape rate through one side of the barrier can be written by

$$\frac{1}{\tau_{th\pm}} = \frac{J_{th\pm}}{qwn}$$

$$= \frac{m^*_{3D}}{m^*_{\|1}} \frac{k_B T}{h} \left(1 \pm \frac{\mu_{3D} F}{k_B T} \sqrt{2m^*_{3D} qV'_{b\pm}}\right) e^{-\frac{qV'_{b\pm} - \varepsilon_1}{k_B T}}. \tag{11}$$

For thin QWs with strong quantum-confinement effect, the derived thermal escape rate should be used instead of the widely-used model $1/\tau_{th\pm} = \sqrt{k_B T / 2\pi m^*_{3D} w^2} \exp[-qV'_{b\pm}/k_B T]$, in which the three-dimensional density of states is assumed for all states.

From Eq. (4), we finally obtain the expression for the *effective thermal mobility*

$$\mu_{th} = \frac{m^*_{3D}}{m^*_{\|1}} \frac{qL^2}{h} \left\{1 + \mu_{3D} \sqrt{\frac{8m^*_{3D} V_b}{qL^2}}\right\} e^{-\frac{q\tilde{V}_b}{k_B T}}, \tag{12}$$

where $q\tilde{V}_b \equiv qV_b - \varepsilon_1$ is the effective barrier height seen by the carriers at the ground state [Fig. 1(b)]. The first term in Eq. (12) corresponds to the so-called thermionic emission from the ground state and the second term corresponds to the field-induced drift transport in the continuum levels. Note that without the approximation (9), our expression for the thermal mobility is reduced to $\mu_{\text{th}} \to \mu_{3D}$ when $V_b \to 0$ and $b \to 0$, as expected for the carrier transport in the smooth bandedge (see Appendix B).

## C. Direct tunneling process

Here, we assume that the tunneling transport is dominated by the sequential (incoherent) tunneling [28] and analyze using the semi-classical approach [35,36]. In this way, the tunneling rate from the $i$th state is estimated from the classical oscillation frequency $f_i$ inside the QW and the WKB tunneling probability $T_i$:

$$\left(\frac{\partial n}{\partial t}\right)_{\text{tun},i} = f_i T_i n_{2D,i}. \quad (13)$$

By using the barrier effective mass $m_b^*$, the perpendicular effective mass $m_{\perp 1}^*$ in the ground state of the QW, and the perpendicular velocity $\sqrt{2\varepsilon_1/m_\perp^*}$, the carrier escape rate by the tunneling process from the ground state (direct tunneling) can be written by

$$\frac{1}{\tau_{\text{tun}\pm}} = \frac{1}{n}\left(\frac{\partial n}{\partial t}\right)_{\text{tun},1}$$
$$= \frac{1}{2w}\sqrt{\frac{2\varepsilon_1}{m_{\perp 1}^*}} e^{-\frac{2}{\hbar}\int_0^b \sqrt{2m_b^*(qV_b-\varepsilon_1 \mp (\frac{1}{2}w+z)qF)}dz}. \quad (14)$$

Therefore, Eqs. (4) and (14) give the *effective tunneling mobility*

$$\mu_{\text{tun}} = \sqrt{\frac{m_b^*}{m_{\perp 1}^*}} \sqrt{\frac{\varepsilon_1}{q\tilde{V}_b}} \frac{qbL^2}{w\hbar} e^{-\frac{2b}{\hbar}\sqrt{2m_b^* q\tilde{V}_b}}. \quad (15)$$

## D. Thermally-assisted tunneling process

We consider the tunneling event from the higher confinement states ($i \geq 2$). From Eqs. (8) and (9b), the density of carriers in the $i$th confinement state is given by $n_{2D,i} = n(m_{\|i}^*/m_{\|1}^*)\exp[-(\varepsilon_i - \varepsilon_1)/k_B T]$. Using Eq. (13) and the $i$th-state perpendicular effective mass $m_{\perp i}^*$, the escape rate by the tunneling from the $i$th state, or thermally-assisted tunneling, can be written as

$$\frac{1}{\tau_{\text{thtun},i\pm}} = \frac{1}{n}\left(\frac{\partial n}{\partial t}\right)_{\text{tun},i}$$
$$= \frac{1}{2w}\frac{m_{\|i}^*}{m_{\|1}^*}\sqrt{\frac{\varepsilon_i}{2m_{\perp i}^*}} e^{-\frac{\varepsilon_i - \varepsilon_1}{k_B T}} \quad (16)$$
$$\times e^{-\frac{2}{\hbar}\int_0^b \sqrt{2m_b^*(qV_b-\varepsilon_i \mp (\frac{1}{2}w+z)qF)}dz}.$$

Using Eq. (4), we obtain the *effective thermally-assisted tunneling mobility*

$$\mu_{\text{thtun}} = \sum_{i\geq 2} \sqrt{\frac{m_b^*}{m_{\perp i}^*}} \frac{m_{\|i}^*}{m_{\|1}^*} \sqrt{\frac{\varepsilon_i}{q\tilde{V}_{b,i}}} \frac{qbL^2}{w\hbar} e^{-\frac{\Delta\varepsilon_i}{k_B T}} e^{-\frac{2b}{\hbar}\sqrt{2m_b^* q\tilde{V}_{b,i}}}, \quad (17)$$

where $\Delta\varepsilon_i \equiv \varepsilon_i - \varepsilon_1$ is the activation energy to the $i$th state and $\tilde{V}_{b,i} \equiv \tilde{V}_b - \Delta\varepsilon_i$ is the effective barrier seen from the carriers in the $i$th state [Fig. 1(b)].

## E. Total effective mobility

Since all the processes above occur in parallel [Fig. 1(b)], the total escape rate can be given by the sum of all the rates $1/\tau_{\text{esc}} = 1/\tau_{\text{th}} + 1/\tau_{\text{tun}} + \sum_{i\geq 2} 1/\tau_{\text{thtun},i}$. Then, it is obvious from Eq. (4) that the total effective mobility is

$$\mu_{\text{eff}} = \mu_{\text{th}} + \mu_{\text{tun}} + \sum_{i\geq 2}\mu_{\text{thtun},i}. \quad (18)$$

In the limit of $V_b \to 0$, there is no confinement states, thus no tunneling event, and the total effective mobility is reduced to $\mu_{\text{eff}} \to \mu_{3D}$ similarly to what we have shown for the effective thermal mobility.

We can obtain the effective mobility of electrons $\mu_{\text{MQW}}^{(e)}$, heavy holes $\mu_{\text{MQW}}^{(hh)}$, and light holes $\mu_{\text{MQW}}^{(lh)}$ by considering the effective masses, bandedges, and energy levels for the corresponding carriers. However, heavy and light holes are not completely independent: they are in the thermal equilibrium sharing the same quasi-Fermi level. From the definition of the hole escape rate, we obtain

$$\frac{1}{\tau_{\text{esc}}^{(h)}} = \frac{1}{n^{(hh)}+n^{(lh)}}\frac{d(n^{(hh)}+n^{(lh)})}{dt}$$
$$= \frac{1}{1+\alpha}\frac{dn^{(hh)}/dt}{n^{(hh)}} + \frac{\alpha}{1+\alpha}\frac{dn^{(lh)}/dt}{n^{(lh)}}$$
$$= \frac{1}{1+\alpha}\frac{1}{\tau_{\text{esc}}^{(hh)}} + \frac{\alpha}{1+\alpha}\frac{1}{\tau_{\text{esc}}^{(lh)}}, \quad (19)$$

where $\alpha$ is the density ratio of the light holes to the heavy holes. By using Eq. (9b), $\alpha$ is given by

$$\alpha = \frac{m_{\parallel 1}^{*(lh)}}{m_{\parallel 1}^{*(hh)}} e^{-\frac{\Delta E_{hh \to lh}}{k_B T}}, \quad (20)$$

where $\Delta E_{hh \to lh}$ is the activation energy for the transition from the heavy-hole ground state to the light-hole ground state. Therefore, we can express the effective hole mobility as

$$\mu_{\text{eff}}^{(h)} = \frac{1}{1+\alpha} \mu_{\text{eff}}^{(hh)} + \frac{\alpha}{1+\alpha} \mu_{\text{eff}}^{(lh)}. \quad (21)$$

This implies that the hole effective mobility is determined from the population-weighted average between the heavy and light hole effective mobilities. It is convenient to define the quantities $\mu_{\text{eff}}^{(p,hh)} \equiv \frac{1}{1+\alpha} \mu_{\text{eff}}^{(hh)}$ and $\mu_{\text{p-eff}}^{(p,lh)} \equiv \frac{\alpha}{1+\alpha} \mu_{\text{eff}}^{(lh)}$ as the *partial effective mobilities* for heavy and light holes, respectively, to quantify the impact from each type of carriers.

Our model provides the formula for the effective mobility which is an explicit function of the material parameters, energy levels, and layer thicknesses of the MQW. Strictly speaking, the various types of effective masses ($m_{3D}^*, m_{\parallel i}^*, m_{\perp i}^*, m_b^*$) and the carrier scattering time ($\langle \tau_{\text{sct}} \rangle$, thus the drift mobility $\mu_{3D}$) are perturbed by the modulated energy profile and are no longer the same as the bulk parameters [8,9,37,38]. In non-ideal MQW structures, the interface roughness between the well and barrier layers can also be an additional scattering path which decreases $\langle \tau_{\text{sct}} \rangle$ [39]. The perturbed parameters should be used in estimating the effective mobility to obtain the precise value. Nevertheless, it is a fair approach to approximate some or all of them with the bulk parameters when the precise values are not available: that is, $m_b^* \approx$ the bulk effective mass in the barrier material, $m_{3D}^*, m_{\parallel i}^*, m_{\perp i}^* \approx m_{\text{well}}^*$ and $\mu_{3D} \approx \mu_{\text{well}}$, where $m_{\text{well}}^*$ is the bulk effective mass and $\mu_{\text{well}}$ is the bulk mobility in the well material. In this way, the effective mobility can be easily estimated once the MQW structure is determined.

## III. EXPERIMENT AND CALCULATION

In this section, the derived effective mobility is discussed and compared with the experimental results. Effective mobilities in MQWs were measured using the optical carrier time-of-flight technique [23]. GaAs p-i-n and n-i-p samples were prepared by metal-organic vapor phase epitaxy under the growth temperature of 610 °C and the pressure of 100 mbar. The sample structure consisted of 200-nm, 600-nm, and 200-nm thick p-region ($2\times10^{18}$ cm$^{-3}$), i-region, and n-region ($1\times10^{17}$ cm$^{-3}$), respectively, with four periods of In$_{0.21}$Ga$_{0.79}$As (5.0 nm)/GaAs$_{0.58}$P$_{0.42}$ (2.0, 5.0, or 8.0 nm) MQW (emission wavelength $\lambda_{\text{em}} = 960$ nm) inserted in the i-region [Fig. 2(a)].

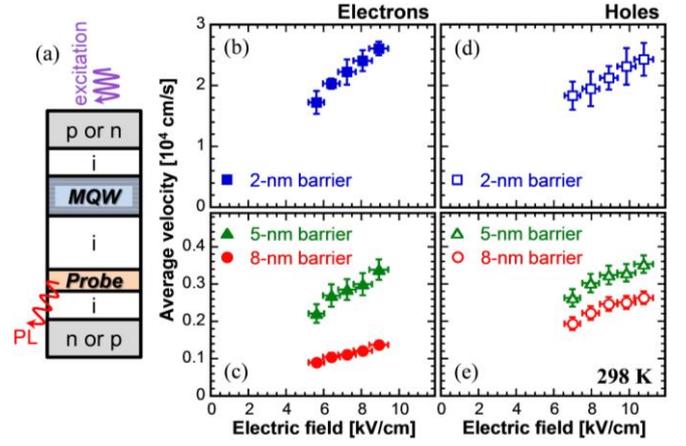

FIG. 2. (a) Schematic of the carrier time-of-flight measurement. Average velocity at 298 K for electrons in the (a) 2-nm, (b) 5-nm, and 8-nm-barrier In$_{0.21}$Ga$_{0.79}$As/GaAs$_{0.58}$P$_{0.42}$ MQWs and (c)-(d) for holes in the similar condition. The error bars for the field and the velocity correspond to the uncertainty of the background carrier concentration and the distribution of the probe response time, respectively.

A 9.5-nm thick In$_{0.25}$Ga$_{0.75}$As single QW ($\lambda_{\text{em}} = 1060$ nm) was inserted underneath each MQW as a probe. The i-GaAs spacers above the MQW and below the probe were designed to be 100 nm thick. The compound compositions and thicknesses of the MQWs and probe QWs were determined by X-ray diffraction. The i-region was slightly n-doped to compensate the p-type background doping incorporated during the growth, confirmed by the Hall measurement to have the net carrier concentration of p-type $(5\pm2)\times10^{14}$ cm$^{-3}$.

The samples were photoexcited by a 405-nm-wavelength pulse laser (repetition rate = 20 MHz, pulse width ≈ 50 ps, pulse energy ≈ 50 pJ) and their time-resolved photoluminescence (TRPL) was detected by a time correlation single-photon counting system (transit time spread ≈ 500 ps). This excitation wavelength has a penetration depth in GaAs less than 20 nm and thus generates carriers only near the sample surface, allowing only electrons to be driven toward the MQW in p-i-n structures and only holes in n-i-p structures. This enables us to extract the transport properties of electrons and holes separately [40]. By measuring the photoluminescence response time of the probe QW, the transport time and the carrier average velocity (the transport time divided by the total MQW thickness) inside the MQW can be obtained. The transport time was calibrated with the reference probe sample where only the MQW was removed (MQW transport time = 0). The electric field across the i-region was controlled by applying the bias voltage in a range of 0.6 to 0.8 V, as the lower bias will result in too weak photoluminescence and the higher bias will cause too strong background electroluminescence from the probe. The electric field value, to a first approximation, was estimated by the simulation on a p-i-n bulk GaAs having the same configuration. The sample temperature was adjusted by a

Peltier temperature controller. More details on the measurement technique have been described in Ref. [23].

Figs. 2(b)-(e) show the average electron and hole velocities at 298 K measured on the p-i-n and n-i-p samples, respectively. We can say that, within a range of electric field in this work (< 12 kV/cm), the average velocity is almost proportional to the electric field, confirming that our experiment condition was in the linear regime for carrier transport. This verifies what we have found from our model in Section III that the velocity-field relation of the sequential transport in MQWs is linear and the $v/F$ ratio converges to a finite value at low field, which is defined as the effective mobility.

The extracted effective mobilities are shown in Fig. 3, together with the calculation results from the model (See Appendix C for the parameters used in the calculation). As can be seen, the calculation results can well predict that the effective mobilities of the MQWs in this study should have values of the order of 0.1-10 cm$^2$/Vs, confirming that our model provides a good approach for the estimation of MQW effective mobilities. There are still some discrepancies between the calculated and the measured values, which are thought to be due to the approximation using the bulk parameters. By breaking down the effective mobilities into the (partial) effective thermal, tunneling, and thermally-assisted tunneling mobilities, the contribution from each carrier escape process can be quantitatively investigated. It is clear that the one-order enhancement of the effective mobilities for both electrons and holes when reducing the GaAsP barrier thickness to 2 nm is due to the direct tunneling process. For MQWs with barriers thicker than 6 nm, the dominant transport is switched from the tunneling to the thermionic emission. The thermal mobility has a small increase as the barriers become thicker since there are less QWs, which act as traps for carriers, per unit length. This is the same reason why the partial effective tunneling mobility of light holes stops increasing when the barrier thickness becomes less than 2 nm. The contribution from the thermally-assisted tunneling process is found to be comparatively small in this structure.

It is important to note that light holes play an important role in the hole transport even when they have small population. For the MQWs in this study, the effective hole mobility is dominated by the transport of light holes, whose density ratio $\alpha$ is estimated to be as low as 0.011. The smaller effective mass in barriers $m_b^*$ and lower effective barrier height $\tilde{V}_b$ (due to the stronger confinement effect) for light holes result in the exponential enhancement of the effective mobility, overwhelming the effect of their small population.

Fig. 4 shows the measured and calculated effective mobilities for higher temperature. In the calculation, we took into account the temperature dependency of the bandedges and the bulk mobilities ($\mu_{3D} \propto T^{-2.3}$ in III-V materials close to room temperature [41]), but kept the other parameters fixed for simplicity. Even though there is a slight overestimate for the hole effective mobility (possibly due to the neglected temperature dependency of some parameters), our model can adequately predict the increase of effective mobility with temperature. With increasing temperature from 298 K to 358 K, the effective thermal mobility of electrons increases significantly whereas the effective tunnel mobility is mostly constant, as can be expected, resulting in the increasing total electron effective mobility by a factor of 2. In contrast, for light holes, the partial effective mobility for the tunneling transport increases with increasing temperature owing to the Boltzmann population factor $\alpha$ of light holes.

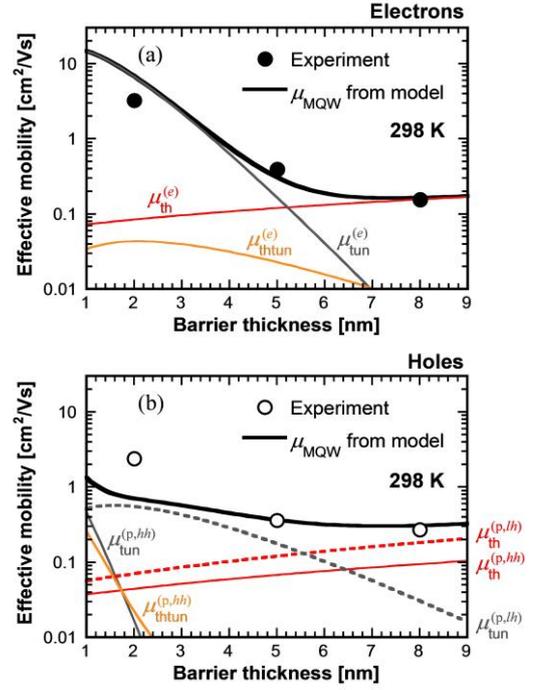

FIG. 3. (a) Electron and (b) hole effective mobilities obtained from the measurement (circles) and the model (lines) at 298 K for different barrier thicknesses. The thick solid lines show the total effective mobilities and the other lines show the contribution of the corresponding processes: the electron effective mobilities, heavy-hole partial effective mobilities, and light-hole partial effective mobilities for the thermionic emission ($\mu_{th}^{(e)}, \mu_{th}^{(p,hh)}, \mu_{th}^{(p,lh)}$), tunneling ($\mu_{tun}^{(e)}, \mu_{tun}^{(p,hh)}, \mu_{tun}^{(p,lh)}$), and thermally-assisted tunneling processes ($\mu_{thtun}^{(e)}, \mu_{thtun}^{(p,hh)}$). Light holes have only the ground state and thus no thermally-assisted tunneling.

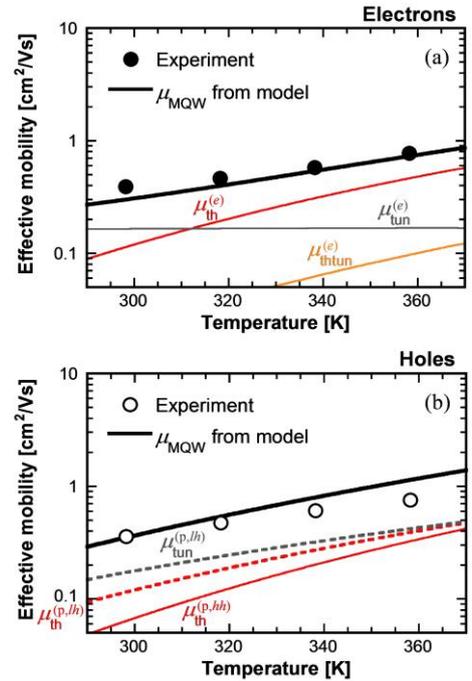

FIG. 4. (a) Electron and (b) hole effective mobilities obtained from the measurement (circles) and the model (lines) in the 5-nm-barrier MQW for different temperatures.

As demonstrated, the availability of the effective mobility model makes it possible to get insight into the behavior of the carrier transport across the MQW in the quantitative manner, particularly for investigating the contribution of each escape process and the impact of each parameter. It can be utilized as a tool for designing the MQW structure to improve the carrier transport and to achieve the target mobility required in each device application.

## IV. EFFECTIVE-MOBILITY APPROXIMATION

The formulae derived in this study allows us to quantify the complicated carrier transport in MQWs with an effective mobility $\mu_{\text{eff}}$. Since $\mu_{\text{eff}}$ has the same dimension as the conventional drift mobility, new approaches to handling MQWs become possible. For instance, in the photodetector/photovoltaic applications, there are sometimes arguments whether bulks with poor carrier transport (e.g. due to the crystal growth issue of thick bulks) or MQWs with high material quality but having potential barriers, are more appropriate for light absorbers with efficient charge extraction [42-45]. In such situations, the concept of effective mobility enables a fair comparison between the carrier transport in MQWs and bulks.

By considering its definition ($v = \mu_{\text{MQW}} F$), it is plausible to say that the similar analogy to the well-established concept of drift mobility can be applied, including the diffusion length $\sqrt{\mu_{\text{MQW}} \tau_r k_B T / q}$ and the conductivity $q n \mu_{\text{MQW}}$. In other words, approximating the MQW structure as an equivalent *quasi-bulk* having effective mobilities, instead of their modulated potential profile, is sufficient to represent the carrier transport behavior in the MQW. The quasi-bulk approach makes the incorporation of MQWs compatible with the conventional drift-diffusion analysis, for example, in the simulation of the p-n junction devices without involving with the additional trap-and-escape mechanism. Note that the effective mobility can only simplify the transport property of the MQW. In order to appropriately simplify the MQW with the quasi-bulk, the additional treatment for other properties such as the density of states have to be considered, which is discussed in Appendix D.

We should keep in mind that the effective-mobility approximation (quasi-bulk approximation) is one approach to deal with the carrier transport complication in MQW structures when the assumptions made above are satisfied. There are some conditions in which this approach may be over-simplified and need further modification as follows.

(1) At low temperature, the carrier thermalization becomes so slow that defining a single quasi-Fermi level $E_F$ in each band can be improper. Furthermore, the dominant transport may be shifted from the sequential to the Bloch transport, which is not taken into account in this study.

(2) At high carrier density, at which carriers are degenerate, the complete expression for the Fermi-Dirac distribution is required in considering the carrier transport.

(3) At high field, the carrier transport becomes non-linear, which may be handled by introducing the field-dependent mobility.

(4) The phonon-limited process, such as phonon-assisted tunneling, may become dominant at low temperature or in some material systems.

Despite some exceptions, the effective-mobility approximation is believed to be a powerful tool when available since most established transport theories in bulks can be directly applied, similarly to how the effective-mass approximation works in successfully describing the carrier transport in bulk crystals near the bandedge.

## V. CONCLUSION

In this study, we investigated the effective carrier mobility for the perpendicular sequential transport in MQWs. The carrier transport is expected, and experimentally confirmed, to be in the linear regime at low field, where the effective mobility can be defined. We derived the expressions of the effective mobilities for the thermionic escape, direct tunneling, thermally-assisted tunneling, and the total effective mobility, which are simple functions of the structure and material parameters. The modeled mobilities were in good agreement with the experimental results, predicting the effective mobilities in $In_{0.21}Ga_{0.79}As/GaAs_{0.58}P_{0.42}$ MQWs to be in the range of 0.1 - 10 cm$^2$/Vs depending on the barrier thickness and the sample temperature. The contribution of light holes was found to be necessary in explaining the hole conduction in spite of their several orders of magnitude smaller population than heavy holes. We suggest that the complicated carrier transport in MQWs at low field, at least to the order of $10^4$ V/cm, can be simplified with virtually free carriers in quasi-bulks with effective mobilities, which can be described with the well-established drift-diffusion transport theory.

## ACKNOWLEDGMENTS

A part of this study is supported by the Research and Development of Ultra-high Efficiency and Low-cost III-V Compound Semiconductor Solar Cell Modules project under the New Energy and Industrial Technology Development Organization, a Grant-in-Aid for JSPS Fellows (15J03447) from the Japan Society for the Promotion of Science, and the National Science Foundation and the Department of Energy under NSF CA No. EEC-1041895.

## APPENDIX A: ASYMPTOTIC EXPRESSION FOR INTEGRATION

Using the asymptotic expansion, we obtain the relation [46]

$$\int_x^\infty e^{-\xi^2} d\xi = \frac{1}{2x} e^{-x^2} \sum_{j=0}^\infty (-1)^j \frac{(2j-1)!!}{(2x^2)^j}. \quad (A1)$$

Then, the integration by parts

$$\int_x^\infty e^{-\xi^2} d\xi = -xe^{-x^2} + 2\int_x^\infty \xi^2 e^{-\xi^2} d\xi \quad (A2)$$

gives the asymptotic expression

$$\int_x^\infty \xi^2 e^{-\xi^2} d\xi = \frac{x}{2} e^{-x^2} \left[ 1 + \sum_{j=0}^\infty (-1)^j \frac{(2j-1)!!}{(2x^2)^{j+1}} \right]$$
$$\approx \frac{x}{2} e^{-x^2} \quad (A3)$$

for $x \gg 1$.

## APPENDIX B: NO-BARRIER LIMIT

We follow the derivation of the effective thermal mobility in Section IIB, but consider in the limit $V_b \to 0$ and $b \to 0$. In this limit, the thermal current in Eq. (7) and the carrier density in Eq. (8) become

$$J_{\text{th}\pm} = \frac{qm_{3D}^* k_B^2 T^2}{2\pi^2 \hbar^3} \left( 1 \pm q \langle \tau_{\text{sct}} \rangle F \sqrt{\frac{\pi}{2m_{3D}^* k_B T}} \right) e^{-\frac{E_0 - E_F}{k_B T}}, \quad (B1)$$

and

$$n = \frac{1}{4} \left( \frac{2m_{3D}^* k_B T}{\pi \hbar^2} \right)^{\frac{3}{2}} e^{-\frac{E_0 - E_F}{k_B T}}, \quad (B2)$$

respectively. Then, the escape rate $1/\tau_{\text{th}\pm} = J_{\text{th}\pm}/qwn$ is given by

$$\frac{1}{\tau_{\text{th}\pm}} = \frac{1}{w} \sqrt{\frac{k_B T}{2\pi m_{3D}^*}} \left( 1 \pm q \langle \tau_{\text{sct}} \rangle F \sqrt{\frac{\pi}{2m_{3D}^* k_B T}} \right). \quad (B3)$$

Therefore, using Eq. (4) and $L = w + b \to w$, we can express the effective thermal mobility as

$$\mu_{\text{th}} = \frac{q \langle \tau_{\text{sct}} \rangle}{m_{3D}^*} = \mu_{3D}. \quad (B4)$$

## APPENDIX C: PARAMETERS FOR EFFECTIVE MOBIILTY CALCULATION

The effective masses, drift mobilities, and band offsets used in the calculation of effective mobilities in Section III are summarized in Table I. We approximated the continuum-state effective masses $m_{3D}^*$, the perpendicular effective masses $m_{\perp i}^*$, and the barrier effective masses $m_b^*$ with the bulk effective masses using Vegard's law. The in-plane effective masses $m_{\|1}^*$ for $In_{0.2}Ga_{0.8}As$ reported in Ref. [47-49] and $m_{\|i}^* \approx m_{\|1}^*$ were used. The continuum-state drift mobilities $\mu_{3D}$ was similarly approximated with the bulk mobilities, with an additional assumption $\mu_{3D}^{(lh)} \approx \mu_{3D}^{(hh)}$ due to the unavailability of the light-hole mobility. The confined energy levels $\varepsilon_i$ were obtained from the envelope-function approximation using the bulk effective masses. The band offset $V_b$ included the effect of strain through the deformation potential. The activation energy for the ground-state light holes $\Delta E_{hh \to lh}$ was $\left( \varepsilon_1^{(lh)} - \varepsilon_1^{(hh)} \right) + \left( E_V^{(hh)} - E_V^{(lh)} \right) = 112$ meV ($E_V$ is the valence bandedge energy).

**Table I**

Material parameters used for the effective mobility calculation at 298 K

| Parameters | Electrons | Heavy holes | Light holes |
|---|---|---|---|
| $m_{3D}^*$, $m_{\perp i}^*$ | $0.058m_0$ | $0.48m_0$ | $0.070m_0$ |
| $m_{\|i}^*$ | $0.069m_0$ | $0.18m_0$ | $0.16m_0$ |
| $m_b^*$ | $0.11m_0$ | $0.65m_0$ | $0.11m_0$ |
| $\mu_{3D}$ [cm$^2$/Vs] | 2000 | 100 | 100 |
| $qV_b$ [eV] | 0.348 | 0.288 | 0.121 |

## APPENDIX D: EQUIVALENT MATERIAL PARAMETERS FOR QUASI-BULK

In order to simplify the MQW ($N$ periods of a QW with thickness $w$ and a barrier with thickness $b$) as the quasi-bulk (homogenous material with total thickness $N(w+b) = NL$) and incorporate it in the drift-diffusion analysis, we need equivalent parameters other than mobilities in order to simulate such equivalent quasi-bulk. We use primes to denote the equivalent parameters in the quasi-bulk. It is obvious that the equivalent bandedges (lowest allowed energies) are

$$E'_C = E_C + \varepsilon_1^{(e)}, \tag{D1a}$$

$$E_V^{\prime(hh)} = E_V^{(hh)} - \varepsilon_1^{(hh)}, \tag{D1b}$$

$$E_V^{\prime(lh)} = E_V^{(lh)} - \varepsilon_1^{(lh)}, \tag{D1c}$$

$$E'_V = \max\left[ E_V^{\prime(hh)}, E_V^{\prime(lh)} \right], \tag{D1d}$$

where $E_C$, $E_V^{(hh)}$, and $E_V^{(lh)}$ are the actual conduction, heavy-hole valence, and light-hole valence bandedges, respectively. (Note that they are denoted as $E_0$ in the main text.) Then, the equivalent bandgap is written by

$$E'_g = E'_C - E'_V. \tag{D2}$$

Since the total number of carriers should remain the same, the equivalent carrier density in the quasi-bulk is

$$n' = \frac{w}{L} n = \frac{m_{\parallel 1}^* k_B T}{L \pi \hbar^2} e^{-\frac{E_0 + \varepsilon_1 - E_F}{k_B T}} \tag{D3}$$

for electrons, heavy holes, and light holes, where $n$ in Eq. (9b) is used. Similarly to the effective density of states in bulks, the equivalent effective density of states, $N'_{C \text{ or } V} \equiv n' \exp\left[ (E'_0 - E_F)/k_B T \right]$, can be expressed by

$$N'_{C \text{ or } V} = \frac{m_{\parallel 1}^* k_B T}{L \pi \hbar^2}. \tag{D4}$$

Then, the equivalent intrinsic carrier concentration $n'_i$, obtained by solving $n'^{(e)} = n'^{(hh)} + n'^{(lh)}$ and $n_i'^2 = n'^{(e)}\left( n'^{(hh)} + n'^{(lh)} \right)$, is

$$n'_i = \sqrt{ N'_C N_V^{\prime(hh)} e^{-E_g^{\prime(e-hh)}/k_B T} + N'_C N_V^{\prime(lh)} e^{-E_g^{\prime(e-lh)}/k_B T} }. \tag{D5}$$

Here, $E_g^{\prime(e-hh)} = E'_C - E_V^{\prime(hh)}$ and $E_g^{\prime(e-lh)} = E'_C - E_V^{\prime(lh)}$ is the equivalent electron-heavy hole and electron-light hole bandgaps, respectively. For degenerate hole bands, Eq. (D5) is reduced to $n'_i = \sqrt{N'_C N'_V} \exp\left[ -E'_g/2 k_B T \right]$ ($N'_V \equiv N_V^{\prime(hh)} + N_V^{\prime(lh)}$), the similar expression to the intrinsic carrier concentration in bulks.

Since the quasi-bulk must have the same total absorption as the actual MQW under the same illumination, the equivalent absorption coefficient has to be given by

$$\alpha' = \frac{\alpha_w w + \alpha_b b}{L}, \tag{D6}$$

where $\alpha_w$ and $\alpha_b$ are the absorption coefficients in the well and barrier layers, respectively. (More complicated expression for $\alpha'$ is required when the multiple reflections inside the MQW is not negligible.) Then, the equivalent parameter for the net radiative recombination coefficient $B_{\text{net}}$ can be obtained from [50-52]

$$B'_{\text{net}} = (1 - P_{\text{abs}}) B'$$
$$= \frac{8\pi (1 - P_{\text{abs}})}{c^2 h^3 n_i^{\prime 2}} \int_{E'_g}^{\infty} \frac{n_r^2(E) \alpha'(E) E^2}{e^{E/k_B T} - 1} dE, \tag{D7}$$

where $c$ is the speed of light, $n_r$ is the diffractive index, and $P_{\text{abs}}$ is the photon-reabsorption probability in the photon recycling process. Note that the coefficient $B_{\text{net}}$ in the actual QW is obtained by substituting $n'_i$ with the actual intrinsic carrier concentration $n_i$ and $\alpha'$ with $\alpha_w$ in Eq. (D7). Since $\alpha_b$ with the high absorption-threshold energy has negligibly small effect on the integration in Eq. (D7), we find the relation

$$B'_{\text{net}} = \frac{L}{w} B_{\text{net}}. \tag{D8}$$

Furthermore, the actual Auger recombination coefficient $C$ in the QW is converted to

$$C' = \left( \frac{L}{w} \right)^2 C \tag{D9}$$

in the quasi-bulk. If the spatial variation of carrier density is sufficiently gradual, it is straightforward to show that

$$\int_0^{NL} R' dz = \int_{\text{QW layers}} R \, dz, \tag{D10}$$

where $R$ represents either the Shockley-Read-Hall, radiative, or Auger recombination. That is, the total carriers, total absorption, and total recombination are invariant in the conversion from the MQW to the quasi-bulk.